\documentclass[11pt,twoside]{article}

\usepackage{graphicx}
\usepackage{asp2006}
\usepackage{epsf}
\usepackage{lscape}

\begin{document}

\title{The All-Sky Catalog of Isolated Galaxies selected from 2MASS}
\author{V. E. Karachentseva}
\affil{Main Astronomical Observatory National Academy of Sciences,
Akad. Zabolotnogo, 27, Kiev, Ukraine}
\author{S. N. Mitronova}
\affil{Special Astrophysical Observatory Russian Academy of Sciences,
Nizhnij Arkhyz, 369167, Russia}
\author{O. V. Melnyk}
\affil{Astronomical Observatory, Kiev National University, 04053, Ukraine}
\affil{Institut d'Astrophysique et de Geophysique, Universit\'{e} de
Li\`{e}ge, All\'{e}e du  6 Ao\^{u}t, 17, B5C, Belgique}
\author{I. D. Karachentsev}
\affil{Special Astrophysical Observatory Russian Academy of Sciences,
Nizhnij Arkhyz, 369167, Russia}

\begin{abstract}

We undertook searches for isolated galaxies based on automatic selection
of infrared sources from the Two Micron All-Sky Survey (2MASS) followed
by a visual inspection of their surrounding. Using a modified criterion
by Karachentseva (1973), we compiled a catalog of 3227 isolated galaxies
(2MIG) containing 6\% of 2MASS Extended Sources (2MASX) brighter than
$K_s = 12^m$ with angular diameters $a_K \geq 30\arcsec$.
  The most isolated 2493 galaxies among them enter in the 2MASS Very
Isolated Galaxy Catalog, 2MVIG. Being situated in the regions of extremely
low mass density, the 2MVIG galaxies can serve as a reference sample for
investigation the influence of environment on structure and evolution of
galaxies.
\end{abstract}

\vspace{-0.5cm}
\section{Introduction}
The importance of studying the isolated galaxies is quite evident now.
A well-defined, numerous, complete and
homogeneous sample of isolated galaxies, which are situated in
low-density regions is necessary
as reference one to study properties of galaxies in groups and clusters.
A special aspect is the investigation of physical processes in isolated
galaxies to explain their star formation rate, chemical abundance,
as well as global morphological properties. A good example of such a
sample is the Catalog of Isolated Galaxies (Karachentseva 1973=CIG)
which contains 1050 galaxies in the northern sky with blue apparent
magnitudes brighter than 15$\fm$7. Here, we present a new catalog of
isolated galaxies, 2MIG, selected from the 2MASX part of of the
$JHK_s$- band 2MASS survey (Jarrett et al. 2000, Skrutskie et al. 1997),
using a modified isolation criterion from CIG.
\section{Selection of isolated galaxies}
The CIG isolation criterion has used the empirically chosen conditions:
\begin{equation}
X_{1i}/a_i\geq s=20,
\end{equation}
\begin{equation}
 4 \geq a_i/a_1 \geq 1/4,
\end{equation}
where indeces ``1'' and ``i'' refer to a fixed galaxy and its neighbours,
respectively.
According to them, the galaxy with a standard angular diameter
$a_1$ is classified as isolated if all its significant neighbours with
their angular diameters $a_i$ locate on the
projected distances not closer than $20a_i$.
Physically, it means  that isolated galaxy did not suffer essential
gravitational influence from nearby galaxies during the last billion
years. To select isolated galaxies from 2MASX we changed the
value "s" in (1) from 20 to 30 because the mean IR galaxy diameter is
1.5 times less than the standard optical one, $a_{25}$. This modified
criterion was applied to 51572  2MASX candidate galaxies with $K_s$
magnitudes being in the interval
\begin{equation}
4\fm0< K_s \leq 12\fm0,
\end{equation}
\begin{equation}
a_K \geq 30\arcsec.
\end{equation}
Hereby, a huge number of 2MASS galaxies with $K_s = 12\fm0 - 14\fm5$,
and $a_K = 10\arcsec - 30 \arcsec$ were taken part in the checking of isolation.
The automatic selection has been performed using the Pleinpot environment
Package developed by the PostgreSQL Global Development Group
for reduction and analysis of astronomical data. As a result we
obtained a sample of 4045 candidate objects, but
250 of them turn out to be non-galaxies:PNs, star clusters and binary
stars. The remaining 3795 were inspected than visually on DSS-1 and DSS-2
surveys using the conditions (1) and (2). The galaxies  without
significant neighbours both in optical and IR bands entered in
the "2MASS Very Isolated Galaxy Catalog" i.e. 2MVIG. There are
2493 of the 2MVIG galaxies, or 4.8\% among 2MASX ones satisfying (3) and
(4). If the significant neighbours have been detected on DSS we checked
their radial velocities using NED and LEDA databases.
When the radial velocity difference was less than 500 km/s we excluded
such candidate galaxy as a member of pair or group; the number of
these galaxies amounts to 568. The remaining 734 galaxies, which have
significant neighbours without measured radial velocities, were added
to the 2MVIG sample as likely isolated ones. These 3227 galaxies
constitute the "2MASS Isolated Galaxies" catalog, 2MIG (Karachentseva
et al. 2010).

\section{The 2MIG catalog}
The data presented in 2MIG are following: (1) -- running number, (2) --
equatorial coordinates (J2000.0) taken from 2MASS, (3) -- galaxy name,
(4) -- angular radius (a large semi-axis) $r_{20fe}$ in arcsec from 2MASS,
(5) -- $K_s$ magnitude from 2MASS, $K_{20fe}$, (6) -- a dimensionless
"separation" between isolated galaxy and its nearest significant neighbour,
$2s=X_{1i}/r_i$, (7) -- heliocentric radial velocity taken from LEDA or NED,
(8) -- morphological type in de Vaucouleurs' scale.
The types were estimated visually on DSS-1, DSS-2 and SDSS surveys,
using also the 2MASS JHK images for central galaxy regions,
(9) -- number of significant neighbours of the isolated galaxy detected
during our additional inspection on DSS. A blank means that the
galaxy belongs to 2MVIG, (10) -- comments about galaxy morphological
peculiarity and its identification in NED with CIG, IRAS as well as the
catalogs and lists of active and peculiar galaxies.

\section{Some general properties of 2MIG}
The cumulative galaxy number versus $K_s$ magnitude is presented in
Fig.1 for different samples.
\begin{figure}[!ht]
\begin{center}
\includegraphics[scale=0.6]{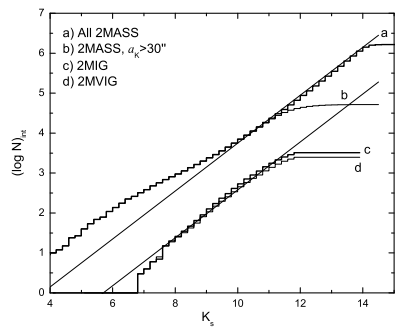}
\end{center}
\caption{The cumulative galaxy number versus $K_s$ magnitude}
\end{figure}
Two parallel lines show the homogeneous disribution with the slope of 0.6 $K_s$.
The isolated 2MIG and 2MVIG galaxies follow well
this distribution except a small lack of bright galaxies
due to the Local supercluster effect. Therefore, the applied criterion
selects about one and the same fraction of isolated galaxies among near as
well as distant galaxies.
\begin{figure}[!h]
\begin{center}
\includegraphics[scale=0.7,angle=-90]{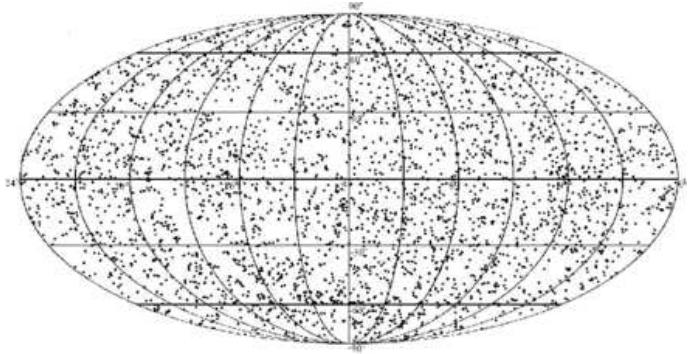}
\end{center}
\caption{Sky distribution of 2MIG galaxies in equatorial coordinates}
\end{figure}
Distribution of 2MIG galaxies on the sky in equatorial coordinates
is given in Fig.2.There is no appreciable enhancement/deficit of isolated galaxies
in the regions of well-known galaxy clusters.

Distribution of 2MIG galaxies on radial velocities is shown in Fig.3.
\begin{figure}[!ht]
\begin{center}
\plottwo{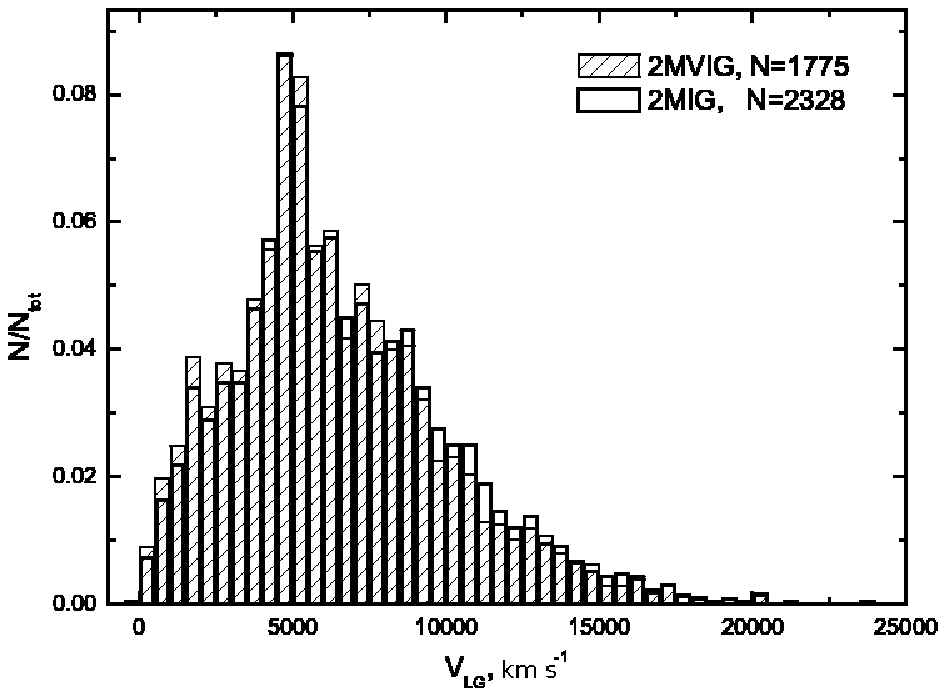}{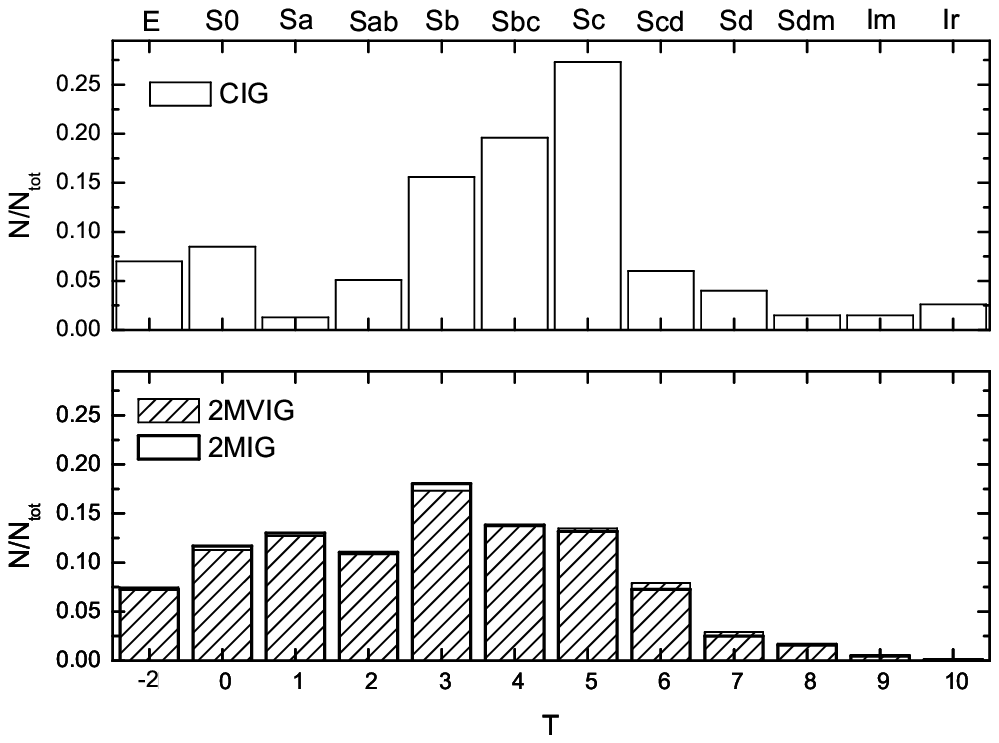}
\end{center}
\caption{Distribution of 2MIG galaxies on their radial velocities, $V_{LG}$
}
\caption{Number rate distribution of CIG, 2MIG and 2MVIG galaxies
}
\end{figure}
The average velocity for them equals to 6570 km/s (or 6360 km/s for
the 2MVIG sample), being very close to the average radial velocity for
CIG galaxies (Verley et al. 2007).

Fig.4 exhibits the number rate distribution of 2MIG and 2MVIG galaxies
by their morphological types.
Elliptical and lenticular galaxies amount to about 19\%, and a fraction
of irregular galaxies does not exceed 1\%.
On the upper panel of Fig.4 we present an appropriate
distribution for CIG galaxies taken from
Hernandes-Toledo et al. (2008). As seen, 2MIG and CIG samples
contain almost the same fraction of early-type galaxies, but
the abundance of the Sa, Sab types is higher in 2MIG in comparison
with CIG. It seems quite expected because of a low ability
of the IR survey to detect blue diffuse objects.

\acknowledgments
This work has been supported by Russian-Ukrainian grant 09-02-90414, Ukrainian-Russian
grant F28.2.059, RFBR grant 07-02-00005 and RFBR-DFG grant 06--02--04017.

{}

\end{document}